\documentclass[12pt]{article}
\pdfoutput=1
\usepackage{subfigure}
\usepackage{amssymb,amsmath}
\usepackage{graphicx}
\usepackage{color}
\usepackage[colorlinks=true
,urlcolor=blue
,citecolor=blue
,linkcolor=blue
,pagecolor=blue
,linktocpage=true
,pdfproducer=medialab
]{hyperref}
\usepackage[a4paper,width=17cm]{geometry}
\makeatletter \renewcommand{\@dotsep}{10000} \makeatother
\usepackage{appendix}

\begin{document}

\begin{center}

 {\Large\bf  The Hydrogen Atom and the Equivalent Form of L{\'e}vy-Leblond Equation
 } \vspace{1cm}

{   Muhammad Adeel Ajaib\footnote{ E-mail: adeel@udel.edu}}

{\baselineskip 20pt \it
Department of Physics, California Polytechnic State University, San Luis Obispo, CA, 93401\\
Department of Mathematics, Statistics and Physics, Qatar  University,  Doha,  Qatar
 } \vspace{.5cm}

{\baselineskip 20pt \it
   } \vspace{.5cm}

\setcounter{footnote}{0}
\vspace{1.5cm}
\end{center}

\begin{abstract}

We discuss the equivalent form of L{\'e}vy-Leblond equation  \cite{LevyLeblond:1967zz, Ajaib:2015uha}  such that the nilpotent matrices are two dimensional. We show that this equation can be obtained in the non-relativistic limit of the (2+1) dimensional Dirac equation.  Furthermore, we analyze the case with four dimensional matrices and propose a Hamiltonian for the equation in (3+1) dimensions and solve it for a Coulomb potential. We show that the quantized energy levels for the hydrogen atom are obtained and the result is consistent with non-relativistic quantum mechanics.

\end{abstract}

\newpage

\section{Introduction}\label{intro}

An equivalent form of the L{\'e}vy-Leblond equation \cite{LevyLeblond:1967zz} was proposed in \cite{Ajaib:2015uha} and it was shown to be consistent with standard quantum mechanical results. The L{\'e}vy-Leblond equation is the analogue of the Dirac equation and describes spin 1/2 particles in the non-relativistic limit. In references \cite{Ajaib:2015uha} and \cite{Ajaib:2015eer} it was shown that the equivalent form of L{\'e}vy-Leblond equation can be employed to solve the step potential problem and the finite potential barrier problem. It was also shown that this equation is the non-relativistic limit of the Dirac equation and the Pauli Hamiltonian can be obtained from this equation by requiring it to be locally invariant. 

In this paper, we present this equation with two dimensional nilpotent matrices and derive it from the (2+1) dimensional Dirac equation. We further illustrate its applications by solving it for a Coulomb potential in (3+1) dimensions when the nilpotent matrices are 4 dimensional. We show that the known expression for the quantized energy levels of the hydrogen atom is obtained from this equation. {The novelty of the approach employed herein is that the spectrum of the hydrogen atom is derived from the L{\'e}vy-Leblond equation which takes into account the spin of the particle in the non-relativistic limit.}

The paper is organized as follows: In section \ref{sec:2dmatrices} we discuss the equivalent form of the L{\'e}vy-Leblond equation in (1+1) dimensions and (2+1) dimensions when the nilpotent matrices are two dimensional. 
In section \ref{sec:hamiltonian} we consider the equation in (3+1) dimensions with 4 dimensional nilpotent matrices and propose a Hamiltonian for this equation. In section \ref{sec:hydrogen} we solve this equation for the Coulomb potential and derive the quantized energy levels of a hydrogen-like atom. We conclude in section \ref{conclude}.

\section{Two Dimensional Matrices}\label{sec:2dmatrices}

In this section we introduce the equivalent form of the equivalent form of the L{\'e}vy-Leblond equation where the nilpotent matrices are 2 dimensional. It was shown in \cite{LevyLeblond:1967zz, Ajaib:2015uha} that the Schr{\"o}dinger equation can be derived from a first order equation similar to the manner in which the Klein Gordon equation can be derived from the Dirac equation. The nilpotent matrices considered in  \cite{LevyLeblond:1967zz, Ajaib:2015uha} were four dimensional. In this section we consider the nilpotent matrices to be 2 dimensional. In (1+1) dimensions the  equivalent form of the L{\'e}vy-Leblond equation is given by \cite{Ajaib:2015uha}
\begin{eqnarray}
-i  \partial_z \psi = (i  \eta \partial_t  + \eta^\dagger m) \psi 
\label{eq:2dmat}
\end{eqnarray}
where the matrix $\eta$ is a 2$\times$2 nilpotent matrix given by
\begin{eqnarray}
\eta = \frac{\sigma_1-i \sigma_2}{\sqrt{2}} = 
{\sqrt{2}}
\left(
\begin{array}{cc}
 0 & 0  \\
 1 & 0  \\
\end{array}
\right)
\label{matrix:eta2d}
\end{eqnarray}
Following the procedure presented in  \cite{Ajaib:2015uha}, we can show that the probability current in this case as well is given by
\begin{eqnarray}
J &=& \psi^\dagger ( \eta+\eta^{\dagger} ) \psi \\
\rho &=&  \psi^\dagger \eta^\dagger \eta \psi
\end{eqnarray}
where $\eta+\eta^{\dagger}=\sqrt{2} \ \sigma_1$ and $\eta^\dagger \eta=I+\sigma_3$. In momentum space, equation (\ref{eq:2dmat}) is given by
\begin{eqnarray}
p_z = (i  \eta \partial_t  + \eta^\dagger m) \psi 
\label{eq:2dmatp}
\end{eqnarray}
The eigenvectors of the momentum operator are given by
\begin{eqnarray}
e_{1,2} &=& 
\left(
\begin{array}{c}
\pm \sqrt{\frac{E}{m}}   \\
 1  \\
\end{array}
\right)
\label{matrix:ev2d}
\end{eqnarray}
which correspond to eigenvalues  $\pm p_z=\pm \sqrt{2Em}$, respectively. Note that, in contrast to the equation with four dimensional matrices \cite{Ajaib:2015uha, Ajaib:2015eer}, the spin of the particle is not taken into account by equation (\ref{eq:2dmat}). The author has checked that the step potential problem and the finite step potential problems solved with equation (\ref{eq:2dmat}) yield results that are consistent with standard quantum mechanical results as in the case of four dimensional matrices \cite{Ajaib:2015uha, Ajaib:2015eer}.  

The (2+1) dimensional version of the L{\'e}vy-Leblond equation for 2$\times$2 matrices, in momentum space, is given by
\begin{eqnarray}
 \mu_i p_i  = (  \eta E + \eta^\dagger m)
\label{eq:2dmat2p1}
\end{eqnarray}
where $\mu_1=I$ and  $\mu_2=i\sigma_3$. 
We can show that equation (\ref{eq:2dmat2p1}) is the non-relativistic limit of the Dirac equation in (2+1) dimensions. Consider the following form of the (2+1) dimensional Dirac equation in momentum space
\begin{eqnarray}
 \gamma_i p_i = ( \sigma_1  E +i\sigma_2 m) \psi 
\label{eq:2dmat2p1de}
\end{eqnarray}
where $\gamma_1=I$ and $\gamma_2=i\sigma_3$. The above equation yields the dispersion relation of a massive relativistic particle in 2D. As in reference \cite{Ajaib:2015eer}, we can substitute $\sigma_1=(\eta+\eta^\dagger)/\sqrt{2}$ and $-i\sigma_2=(\eta-\eta^\dagger)/\sqrt{2}$ and apply the non-relativistic limit  $E-m \simeq E'$ and $E+m \simeq 2m$ to obtain equation (\ref{eq:2dmat2p1}) from (\ref{eq:2dmat2p1de}).

Note also that in the limit $m=0$, equation (\ref{eq:2dmat2p1de}) reduces to the Dirac equation for massless fermions
\begin{eqnarray}
E=\sigma_i p_i
\end{eqnarray}
which, as an example, is employed to describe massless fermions in condensed matter systems such as graphene.

\section{The Hamiltonian}\label{sec:hamiltonian}

In this section we present the Hamiltonian corresponding to the equivalent form of the L{\'e}vy-Leblond equation with four dimensional matrices and discuss the constants of motion.  The (3+1) dimensional version of equation is given by \cite{Ajaib:2015uha, Ajaib:2015eer}
\begin{eqnarray}
-i \gamma_i \partial_i \psi = (i  \eta \partial_t  + \eta^\dagger m) \psi 
\label{mse-eq-3d}
\end{eqnarray}
where $\gamma_i$ are the Dirac gamma matrices and $\eta=(\gamma_0+i \gamma_5)/\sqrt{2}$. One of the issues in obtaining the Hamiltonian of equation (\ref{mse-eq-3d}) is that the matrix $\eta$ is singular.  Recently, a Hamiltonian was proposed in \cite{Sobhani:2016xao} and we adopt a different approach herein. In order to obtain the Hamiltonian we replace $\eta \rightarrow \eta^\prime=\eta-\epsilon \eta^\dagger$ and analyze the limit $\epsilon \rightarrow 0$. We thereby obtain the following Hamiltonian for equation (\ref{mse-eq-3d})
\begin{eqnarray}
H=\eta^{\prime -1} (-i \gamma_i \partial_i-m \eta^{\prime \dagger})
\label{eq:hamiltonian}
\end{eqnarray}
where $\eta^\prime=\eta-\epsilon \eta^\dagger$ and we choose $\hbar=c=1$. In the limit $\epsilon \rightarrow 0$ two of the eigenvalues of the Hamiltonian in (\ref{eq:hamiltonian}) are finite where as two approach infinity
\begin{eqnarray}
E_{1,2} &=& \frac{\vec{p}^2}{2m} \\
E_{3,4} &=& -\frac{\vec{p}^2}{2m}+\frac{m}{\epsilon}
\end{eqnarray}
The Hamiltonian yields the two finite energy states in addition to negative energy states with an infinite part. The infinity associated with the negative energy states can be interpreted as the ``sea" of filled negative energy states. For the negative energy states we can define the renormalized energy as 
\begin{eqnarray*}
E^\prime_{3,4}=E_{3,4}-\frac{m}{\epsilon}= -\frac{\vec{p}^2}{2m} 
\end{eqnarray*}
The Hamiltonian (\ref{eq:hamiltonian}) is not hermitian however the eigenvalues of the operator are real. 
Interestingly, the Hamiltonian (\ref{eq:hamiltonian}) commutes with the total angular momentum operator $\vec{J}=\vec{L}+1/2 \vec{\Sigma}$ and the operators $J^2, \ J_z$ and $K$, i.e.,
\begin{eqnarray*}
[H,J^2] &=& 0 \\
\left[H,J_z\right] &=& 0 \\
\left[H,K\right] &=& 0 
\end{eqnarray*}
and the operator $K$ also commutes with the total angular momentum operators $J^2$ and $J_z$. The operator $K$ is given by 
\begin{eqnarray}
K &=& i \gamma_5 \gamma_0 (\vec{\Sigma}.\vec{J}-\frac{1}{2} I ) \\
 &=& i \gamma_5 \gamma_0 (\vec{\Sigma}.\vec{L}+ I ) \nonumber \\
 &=& 
i\left(
\begin{array}{cc}
 0 &  \vec{\sigma}.\vec{L}+I \\
 -\vec{\sigma}.\vec{L}-I &  0
\end{array}
\right)
\label{matrix-k}
\end{eqnarray}
where $\vec{J}=\vec{L}+1/2 \vec{\Sigma}$. We can construct simultaneous eigenfunctions of the mutually commuting operators $H$, $J^2$, $J_z$ and $K$. The corresponding eigenvalues of these operators are denoted by $E$, $j(j+1)$, $m_j$ and $-\kappa$. We consider the following four component wave function as the simultaneous eigenfunction of these operators 

\begin{eqnarray}
\psi=
\left(
\begin{array}{c}
 \psi_A \\
 \psi_B
\end{array}
\right)
=
\left(
\begin{array}{c}
  g(r) Y^{j,m_j}_{l_A}(\theta, \phi)  \\
 i f(r) Y^{j,m_j}_{l_B}(\theta, \phi)
\end{array}
\right)
\equiv
\left(
\begin{array}{c}
  g(r) Y_{A}  \\
 i f(r) Y_{B}
\end{array}
\right)
\label{matrix-2}
\end{eqnarray}
and for the angular part $Y^{j,m_j}_{l_A, l_B}(\theta, \phi)$ we consider the case $\theta=0$ \cite{Sakurai:2014}
\begin{eqnarray}
Y^{j,m_j}_{l=j\mp 1/2}(\theta=0,\phi)=\sqrt{\frac{j+1/2}{4\pi}} \left(
\begin{array}{c}
  \pm \delta_{m,1/2} \\
 \delta_{m,-1/2}
\end{array}
\right)
\label{eq:angular-part}
\end{eqnarray}
where $l_A=j+1/2$ and $l_B=j-1/2$. {We choose $\theta=0$ because the effect of the pseudo-scalar operator $\vec{\sigma}.\vec{r}/r$ on $Y^{j,m_j}_{l}$ is independent of $\theta$ \cite{Sakurai:2014}}. The eigenvalues of the operator $K$ are given by
\begin{eqnarray}
K \psi &=& - \kappa \psi
\end{eqnarray}
Since $J^2=K^2-1/4 I$, the eigenvalues of the two operators are related as $\kappa=\pm(j+1/2)$. Plugging in for $K$ yields the following equations
\begin{eqnarray}
\vec{\sigma}.\vec{L} \psi_A &=& - i \kappa \psi_B-\psi_A \label{eq:sla}\\
\vec{\sigma}.\vec{L} \psi_B &=&  i \kappa \psi_A-\psi_B \label{eq:slb}
\end{eqnarray}
In addition we have the following eigenvalue equations
\begin{eqnarray}
\vec{J}^2 \psi_{A,B} &=& j(j+1)  \psi_{A,B} \\
J_z \psi_{A,B} &=& j_z  \psi_{A,B}
\end{eqnarray}

\section{Solution for the Coulomb Potential and the Hydrogen-like Atom}\label{sec:hydrogen}

In this section, we study the problem of an electron bound to a nucleus by a Coulomb potential for a hydrogen-like atom (For the analysis with the Dirac equation and further details the reader is referred to \cite{Sakurai:2014, Sakurai:1967, Greiner:2000}). For the case of a Coulomb potential, the Hamiltonian is given by
\begin{eqnarray}
H=\eta^{\prime -1} (-i \gamma_i \partial_i-m \eta^{\prime \dagger})+ V(r)
\end{eqnarray}
where $V(r)=-Z\alpha/r$, $\alpha\approx 1/137$ is the fine structure constant and $Z$ is the atomic number of the atom. Since $\psi$ is an eigenstate of the Hamiltonian
\begin{eqnarray*}
H \psi &=& E \psi \\
\eta^{\prime -1} (-i \gamma_i \partial_i-m \eta^{\prime \dagger}) \psi +V(r) \psi &=& E \psi \\
( \gamma_i p_i-m \eta^{\prime \dagger})\psi &=& (E-V(r))\eta^{\prime} \psi
\end{eqnarray*}
where $p_i=-i \partial_i$. We therefore obtain
\begin{eqnarray}
 \gamma_i p_i \psi = (  \eta^\prime (E-V(r))  +  \eta^{\prime \dagger} m) \psi 
\label{eq-3d}
\end{eqnarray}
\begin{eqnarray}
\vec{\sigma}.\vec{p}
\left(
\begin{array}{c}
  \psi_B \\
 -\psi_A
\end{array}
\right)
= \frac{1}{\sqrt{2}}
\left(
\begin{array}{cc}
 a^\prime (E-V+m) &  i a (E-V-m) \\
 i a (E-V-m) &  -a^\prime (E-V+m)
\end{array}
\right)
\left(
\begin{array}{c}
  \psi_A \\
 \psi_B
\end{array}
\right)
\label{matrix-2b}
\end{eqnarray}
where $a^\prime=1-\epsilon$ and $a=1+\epsilon$. For brevity, we write
\begin{eqnarray}
\vec{\sigma}.\vec{p}
\left(
\begin{array}{c}
  \psi_B \\
 -\psi_A
\end{array}
\right)
=
\left(
\begin{array}{cc}
 h_1 &  i h_2 \\
 i h_2 &  -h_1
\end{array}
\right)
\left(
\begin{array}{c}
  \psi_A \\
 \psi_B
\end{array}
\right)
\label{matrix-3}
\end{eqnarray}
where
\begin{eqnarray}
h_1(r)=a^\prime/\sqrt{2} (E-V(r)+m) \label{eq:h1} \\
h_2(r)=a/\sqrt{2} (E-V(r)-m) \label{eq:h2}
\end{eqnarray}
The operator $\vec{\sigma}.\vec{p}$ can be written in terms of the radial and angular operators as 
\begin{eqnarray}
\vec{\sigma}.\vec{p} = \frac{1}{r} \frac{\vec{\sigma}.\vec{r}}{r}\left( -i r \frac{\partial}{\partial r} + i \vec{\sigma}.\vec{L} \right)
\label{eq:sigdp}
\end{eqnarray}
The operator $\vec{\sigma}.\vec{r}/r$ is a pseudo scalar and changes the parity of the state, i.e.
\begin{eqnarray}
\frac{\vec{\sigma}.\vec{r}}{r} Y_A=-Y_B
\label{eq:paritych}
\end{eqnarray}
with $(\vec{\sigma}.\vec{r}/{r})^2=1$. We are interested in the effect of the operator $\vec{\sigma}.\vec{r}/r$ on $Y^{j,m_j}_{l}$ and due to its pseudo-scalar nature its effect on $Y^{j,m_j}_{l}$ is independent of $\theta$ \cite{Sakurai:2014}. So we choose $\theta=0$ for the angular part and employ the expression given in equation (\ref{eq:angular-part}) for the analysis. Plugging (\ref{eq:sigdp}) in (\ref{matrix-3}) we obtain the following two equations
\begin{eqnarray}
\frac{1}{r} \frac{\vec{\sigma}.\vec{r}}{r}\left( -i r \frac{\partial}{\partial r} + i \vec{\sigma}.\vec{L} \right) \psi_B = h_1 \psi_A +i h_2 \psi_B \\
-\frac{1}{r} \frac{\vec{\sigma}.\vec{r}}{r}\left( -i r \frac{\partial}{\partial r} + i \vec{\sigma}.\vec{L} \right) \psi_A = i h_2 \psi_A - h_1 \psi_B
\end{eqnarray}
Plugging in $\psi_A=g(r) Y_{A}$ and $\psi_B= i f(r) Y_{B}$ and using  equations (\ref{eq:angular-part}), (\ref{eq:sla}), (\ref{eq:slb}), and (\ref{eq:paritych})  results in the following equations	
\begin{eqnarray}
\frac{\partial f}{\partial r}+\frac{1}{r} f+h_2 f+h_1 g +\frac{\kappa}{r} g =0 \\
\frac{\partial g}{\partial r}+\frac{1}{r} g-h_2 g-h_1 f +\frac{\kappa}{r} f =0
\end{eqnarray}
The above equations are obtained for the $m=+1/2$ case. The analysis below also holds for the $m=-1/2$  which yields the similar results. Next, plugging in $f(r)=F(r)/r$ and $g(r)=G(r)/r$ and using (\ref{eq:h1}) and (\ref{eq:h2}) we obtain the following equations
\begin{eqnarray}
\frac{\partial F}{\partial r}+\left( q_1 + \frac{q_2}{r} \right) F+\left( p_1 + \frac{p_2}{r} +\frac{\kappa }{r} \right) G =0
\label{eq:Fr} \\
\frac{\partial G}{\partial r}-\left( q_1 + \frac{q_2}{r} \right) G+\left(-p_1 - \frac{p_2}{r} +\frac{\kappa }{r} \right) F =0
\label{eq:Gr}
\end{eqnarray}
Here we have defined the following constants
\begin{eqnarray}
p_1=\frac{a}{\sqrt{2}} (E+m), \ \
p_2=\frac{a}{\sqrt{2}} {Z\alpha} \\
q_1=\frac{a^\prime}{\sqrt{2}} (E-m), \ \
q_2=\frac{a^\prime}{\sqrt{2}} {Z\alpha}
\end{eqnarray}
We postulate series solutions of (\ref{eq:Fr}) and (\ref{eq:Gr}) of the form
\begin{eqnarray}
F(r)=e^{-\lambda r}  \sum\limits_{n=0}^{\infty}a_n r^{s+n} \label{eq:series-1}\\
G(r)=e^{-\lambda r}  \sum\limits_{n=0}^{\infty}b_n r^{s+n}
\label{eq:series-2}
\end{eqnarray}
Plugging (\ref{eq:series-1}) and (\ref{eq:series-2}) in (\ref{eq:Fr}) and (\ref{eq:Gr}) we obtain the following equations for the coefficients of the two series
\begin{eqnarray}
 q_2  a_{n + 1} + (n + 1)  a_{n+1} + q_1  a_{n} + 
  s a_{n+1} - \lambda a_{n} + p_2 b_{n+1} + \kappa  b_{n+1} + p_1  b_{n}=0
  \label{eq:rec-1} \\
-p_2  a_{n+1} + \kappa   a_{n+1} - p_1  a_{n} - 
  q_2  b_{n+1} + (n + 1)  b_{n+1} - q_1  b_{n} + 
  s b_{n+1} - \lambda b_{n}=0
    \label{eq:rec-2} 
\end{eqnarray}
For $n=-1$, the above equations are given as follows
\begin{eqnarray}
q_1 a_{-1} + (q_2 + s) a_{0} + 
  p_1 b_{-1} + (p_2 + \kappa) b_{0} &=& \lambda a_{-1} \\
p_1 a_{-1} + p_2 a_{0} + (q_1 + \lambda) b_{-1} + q_2 b_{0} &=&
 \kappa  a_{0} + s b_{0}
\end{eqnarray}
Setting $a_{-1}=b_{-1}=0$ yields
\begin{eqnarray}
(q_2 + s) a_{0} +  (p_2 + k) b_{0} &=& 0 \\
  p_2 a_{0} + q_2 b_{0} &=&  \kappa  a_{0} + s b_{0}
\end{eqnarray}
The solution of the above equations is
\begin{eqnarray}
s &=& \pm \sqrt{\kappa^2+q_2^2-p_2^2 }
\end{eqnarray}
For the wave function to be normalizable we choose the positive sign of the square root. Furthermore, the series of $F(r)$ and $G(r)$ must terminate at some $n=n^{\prime}$ for the state to be normalizable. This implies that the coefficients $a_i=b_i=0$ for $i=n^\prime+1$ and we obtain the following relation
\begin{eqnarray}
b_{n^\prime}=\frac{\sqrt{q_1^2-p_1^2}-q_1}{p_1} a_{n^\prime}
\label{eq:bnp}
\end{eqnarray}
where we have chosen $\lambda=\sqrt{q_1^2-p_1^2}$. Next we solve the recursion relations (\ref{eq:rec-1}) and (\ref{eq:rec-2})  for $n=n^\prime-1$
\begin{eqnarray}
(q_1 - \lambda) a_{n^\prime-1} + (q_2 + n^\prime + s) a_{n^\prime} + 
 p_1 b_{n^\prime-1} + (p_2 + k) b_{n^\prime}=0 \label{eq:np1}\\
-p_1 a_{n^\prime-1} + (-p_2 + k) a_{
   n^\prime} - (q_1 + \lambda) b_{n^\prime-1} + (-q_2 + n^\prime + s) b_{n^\prime}=0  \label{eq:np2}
\end{eqnarray}

\noindent
Multiplying (\ref{eq:np1}) by $1/(\lambda-q_1)$ and (\ref{eq:np2}) by $1/p_1$ and subtracting we obtain the following equation

\begin{eqnarray}
p_1((q_2 + n^\prime + s) a_{n^\prime} + (p_2 + k) b_{n^\prime})) + (
  q_1 - \lambda)((-p_2 + k) a_{n^\prime} + (-q_2 + n^\prime + s) b_{n^\prime})=0
\end{eqnarray}
Taking the limit $\epsilon \rightarrow 0$ ($a=a^\prime=1$, $s=\kappa=j+1/2$) and using equation (\ref{eq:bnp}) and $\lambda=\sqrt{-2 E m}$ we obtain the relation for the energy level
\begin{eqnarray}
E = -\frac{m Z^2 \alpha^2}{2 n^2}
\label{eq:energy-level}
\end{eqnarray}
where $n=n^\prime + s=n^\prime+j+1/2=n^\prime+l+1$ 
 is the principal quantum number. The above equation is the known expression for the energy level of a hydrogen-like atom. For the hydrogen atom $Z=1$. Note that the parameter $s$ has to be positive and since $s=\kappa$, only $\kappa=+(j+1/2)$ is relevant. The functions $f(r)$ and $g(r)$ are therefore given by
\begin{eqnarray}
f(r)=e^{-\sqrt{-2Em}  r} \ r^{\kappa-1}  \sum\limits_{m=0}^{\infty}a_m r^{m}
= e^{-\frac{m Z \alpha}{n} r} \ r^{l}  \sum\limits_{m=0}^{\infty}a_m r^{m}
 \label{eq:series-3}\\
g(r)=e^{-\sqrt{-2Em}  r} \ r^{\kappa-1} \sum\limits_{m=0}^{\infty}b_m r^{s+m}
= e^{-\frac{m Z \alpha}{n} r} \ r^{l}  \sum\limits_{m=0}^{\infty}b_m r^{m}
\label{eq:series-4}
\end{eqnarray}
The ground state wave function ($n^\prime=0, \ \kappa=1, \ j=1/2$) of the Hydrogen atom can be written as 
\begin{eqnarray}
\psi_{gd}=
N \frac{1}{\sqrt{4\pi}} e^{-Z r/a_{B}}\left(
\begin{array}{c}
  g(r) \chi_s  \\
 -i f(r) \  {\vec{\sigma}}.{\hat{{r}}} \ \chi_s
\end{array}
\right)
\label{matrix-3b}
\end{eqnarray}
where $a_B=1/\alpha m$ is the Bohr's radius and
\begin{eqnarray}
\vec{\sigma}.\hat{r}
=
\left(
\begin{array}{cc}
 \cos\theta &  e^{-i\phi} \sin\theta \\
e^{i\phi} \sin\theta &  -\cos\theta
\end{array}
\right)
\end{eqnarray}
\begin{eqnarray}
\chi_s= 
\left(
\begin{array}{c}
 1 \\
 0 \\
\end{array}
\right) \, 
\mathrm{or}
\,
\left(
\begin{array}{c}
 0\\
 1 \\
\end{array}
\right)
\label{matrix-5}
\end{eqnarray}
for the spin quantum number $m_s=+1/2$ and $m_s=-1/2$. For $m_s=+1/2$ the wave function is given by
\begin{eqnarray}
\psi_{gd}= N \frac{1}{\sqrt{4\pi}} e^{-Z r/a_{B}}
\left(
\begin{array}{c}
 1 \\
 0 \\
 -i d_0 \cos\theta \\ 
 -i d_0 \sin\theta e^{i \phi}\\
\end{array}
\right)
\label{matrix-4}
\end{eqnarray}
For $m_s=-1/2$
\begin{eqnarray}
\psi_{gd}= N \frac{1}{\sqrt{4\pi}} e^{-Z r/a_{B}}
\left(
\begin{array}{c}
 0 \\
 1 \\
 -i d_0 \sin\theta e^{-i \phi} \\ 
 i d_0 \cos\theta \\
\end{array}
\right)
\label{matrix-4b}
\end{eqnarray}
where $d_0=a_0/b_0=\frac{2-\sqrt{2} Z \alpha}{2+\sqrt{2} Z \alpha}$. The normalization constant is given by
\begin{eqnarray}
N=2\sqrt{\pi}\left(\frac{Z}{a_B}\right)^{3/2} \frac{2+\sqrt{2}{Z \alpha}}{\sqrt{2+Z^2\alpha^2}}
\end{eqnarray}

\section{Conclusion} \label{conclude}

We presented the equivalent form of the L{\'e}vy-Leblond equation with two dimensional nilpotent matrices and showed that in (2+1) dimensions it can be obtained from the Dirac equation in the non-relativistic limit. In (3+1) dimensions we also proposed a Hamiltonian for this equation with four dimensional nilpotent matrices and showed that the quantized energy level of the hydrogen atom are obtained when the equation is solved for a Coulomb potential. We also derived the ground state wave function for spin up and down electron for a hydrogen-like atom. {The novelty of this approach is that the spin of the electron is taken into account in the non-relativistic limit to obtain the spectrum of the hydrogen atom}. This analysis further illustrates the application of this equation which allows for additional insights into a problem corresponding to the spin of the particle.

\section{Acknowledgments}
The author would like to thank Fariha Nasir, Warren Siegel and Mansoor Ur Rehman  for useful discussions and suggestions. The author is also grateful to Amer Iqbal for bringing reference \cite{LevyLeblond:1967zz} to the authors attention.

\end{document}